\documentclass[aps,showpacs,preprintnumbers,amsmath,amssymb]{revtex4}
\usepackage[english]{babel}
\usepackage[latin1]{inputenc} 
\usepackage{color}
\usepackage{graphicx}
\usepackage{graphics}

\begin{document}
%
%
\title{Velocity, temperature and normal force dependence on friction:\\
An analytical and molecular dynamic study}
\author{R.A. Dias}
\email{radias@fisica.ufmg.br} \affiliation{Departamento de
F\'{\i}sica - ICEX - UFMG 30123-970 Belo Horizonte - MG, Brazil}
\author{P. Z. Coura} \email{pablo@fisica.ufjf.br}
\affiliation{Departamento de F\'{\i}sica - ICE - UFJF  Juiz de
Fora - MG, Brazil}
\author{M. Rapini} \email{mrapini@fisica.ufmg.br}
\affiliation{Departamento de F\'{\i}sica - ICEX - UFMG  30123-970
Belo Horizonte - MG, Brazil}
\author{B.V. Costa} \email{bvc@fisica.ufmg.br}
\affiliation{Laborat\'orio de Simula\c{c}\~ao - Departamento de
F\'{\i}sica - ICEX - UFMG 30123-970 Belo Horizonte - MG, Brazil}
\date{\today}
\begin{abstract}
In this work we propose an extension to the analytical
one-dimensional model proposed by E. Gnecco (Phys. Rev. Lett.
84:1172) to describe friction. Our model includes normal forces and
the dependence with the angular direction of movement in which the
object is dragged over a surface. The presence of the normal force
in the model allow us to define judiciously the friction
coefficient, instead of introducing it as an {\sl a posteriori}
concept. We compare the analytical results with molecular dynamics
simulations. The simulated model corresponds to a tip sliding over a
surface. The tip is simulated as a single particle interacting with
a surface through a Lennard-Jones $(6-12)$ potential. The surface is
considered as consisting of a regular BCC(001) arrangement of
particles interacting with each other through a Lennard-Jones
$(6-12)$ potential. We investigate the system under several
conditions of velocity, temperature and normal forces. Our
analytical results are in very good agreement with those obtained by
the simulations and with experimental results from E. Riedo (Phys.
Rev. Lett. 91:084502) and Eui-Sung Yoon (Wear 259:1424-1431) as
well.
\end{abstract}
\pacs{46.55.+d, 07.79.Lh, 07.79.Sp, 81.40.Pq, 62.20.Qp, 68.35.Af}
\maketitle
%
%
\section{Introduction\label{Intro}}
Understanding the origin of tribological phenomena is a fascinating
and challenging enterprize. The classical point of view of the
frictional phenomena, can be synthesized in the three laws of
friction, valid in the macroscopic scale
\cite{Nanoscience,HBMicroNanoTribo}:
\begin{enumerate}
\item Friction is independent of the apparent area of contact,
\item Friction is proportional to the applied load. The ratio between
the friction force and the applied load is named the coefficient of
friction ($\mu=f_L/f_N$) and it is larger for static friction than
for kinetic friction,
\item Kinetic friction is independent of the relative sliding
velocity.
\end{enumerate}
Since new tools, such as the atomic force microscopy (AFM), have
made possible to examine the friction phenomenon in great detail
these laws have been questioned in systems with dimensions
approaching the nanometer scale. At the same time, the development
of ultra fast computers have allowed to test new theories on the
nano-scale friction world. Although tribology is an old science, and
in spite of the efforts and progress made by scientists and
engineers in the last years, tribology is still far from being a
well-understood subject. In fact, it is incredible that even knowing
several properties as surface energy, elastic properties and loss
properties, a friction coefficient cannot be found by using an {\sl
a priori} calculation. Although in the macroscopic scale the
friction force, $f_L$, is independent of the relative velocity, in
the nanometric scale some authors
\cite{Gnecco,Riedo,EGneccoWear254,chen:236102} observed that the
mean value of the friction force presents a logarithmic velocity
dependence. Another important result was the conclusion that
friction force is proportional to the effective contact area down to
the nanometer scale \cite{EuiSungYoon}. An analytical
one-dimensional model known as Tomlison model \cite{Nanoscience} was
able to explain several features of the nanoscopic friction. Using
the Tomlinson model in the limit of low velocities Gneco et al
\cite{Gnecco} showed that the friction force has a logarithmic
dependence with the velocity. Using the same ideas, but in the limit
of higher velocities, Sang et al. \cite{Sang} obtained that the
friction force is proportional to $|\ln(v)|^{2/3}$, were $v$ is the
relative velocity. Using a first principle model Persson
\cite{PerssonWear} was able to show that in the limit of small
contact areas the result of Sang is recovered while in the limit of
large contact areas the Gneco result fitted better. The aim in this
work is to develop a model from first principle by extending the
one-dimensional model proposed by Sang et al. \cite{Sang} to
three-dimensions. Based in our approach we obtain a friction
coefficient which can be calculated knowing simple parameters of the
model (As bound energies and the positions of the minima between
atoms.). Such parameters can be obtained by using ab-initio
calculations or measured experimentally by using FFM (Friction force
microscopy) \cite{Mate,Gnecco,EGneccoWear254,ASocoliuc} or DFS
(Dynamic force microscopy) \cite{PRB62}. We study the sliding
frictional process by using two approaches:
\begin{itemize}
\item Developing an analytic model that considers the potential
energy between a atom in the tip and the surface atoms described by
the model presented by W. A. Steele \cite{Steele} as a sum of
pair-wise $(6-12)$ LJ potentials. The analytic treatment extends the
one dimensional model proposed by Riedo et al.\cite{Riedo} including
the normal force and as a consequence, the effects of adhesion
energies.
\item Using MD simulations by considering that the potential
energy between the tip's atom and the surface atoms is described as
a sum of $(6-12)$ LJ potentials (See ref. \cite{RadiasBJP} and
references there in.).
\end{itemize}
%
%
\section{First principle model\label{FirstPricMod}}
\vspace{0.1cm}%
\begin{figure}[htbp]
  \includegraphics[width=9.0cm,height=4.0cm]{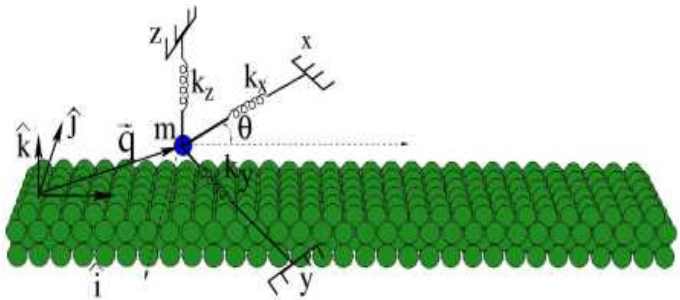}\\
  \caption{\scriptsize{Perspective view of a typical
  initial configuration [BCC(001) geometry] for the FFM experiment.}}\label{FFMesqSim}
\end{figure}
\begin{figure}[htbp]
\begin{center}
\includegraphics[height=5.0cm,width=6.0cm]{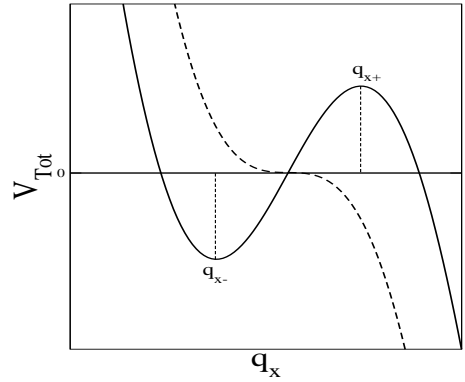}
\end{center}\vspace*{-0.4cm}
\caption{{Schematic figure illustrating the critical state for two
different positions of the cantilever, $x<x_c$ (full line) and
$x=x_c$ (dashed line). The marked points define the energy barrier
that prevents a jump from one stable position ($q_{x-}$) to another
(equation \ref{DefDeltaE}). Figure from reference \cite{Riedo}. }}
\label{Vtot-qx-Barreiraq+q-}
\end{figure}

In this section we show a general picture of the nature of the
kinetic friction. Consider the sliding system shown in figure
(\ref{FFMesqSim}) in which one particle of mass $m$ is connected
through a spring to a cantilever or drive. The particle experiences
a total force described by the potential \cite{Nanoscience}
\begin{eqnarray}\label{PotUTot}
V_{Tot}=\frac{1}{2}\left[ (\vec{q}-\vec{r}) \cdot
\overleftrightarrow{k} \cdot (\vec{q}-\vec{r}) \right]+
V_{int}(\vec{q}),
\end{eqnarray}
where
\begin{equation}\label{SpringTensor}
\overleftrightarrow{k}= \left[
\begin{array}{ccc}
k_x & 0   & 0 \\
0   & k_y & 0 \\
0   & 0   & k_z \\
\end{array}
\right]
\end{equation}
represents the harmonic spring constant complying the cantilever
with the tip, $V_{int}(\vec{q})$ is the surface-tip corrugated
potential, $\vec{q}=(q_x,q_y,q_z)$ are the coordinates of the tip
and $\vec{r}=(x,y,z)$ the coordinates of the support.
As we are interested to study the influence of the normal force on
the sliding process, let us first note that the critical state
(Where we denote the drive position by $\vec{r}_{c}=(x_c,y_c,z_c)$
and the particle position by $\vec{q}_{c}=(q_{xc},q_{yc},q_{zc})$)
is the position where the tip jumps from a stable position in the
surface to the next one. To illustrate the occurrence of the
critical state, we show in the figure \ref{Vtot-qx-Barreiraq+q-} the
total potential energy of the tip as a function of $q_x$, for two
different positions of the cantilever, $x<x_c$ (full line) and
$x=x_c$ (dashed line). As a matter of clarity we restrict this
figure to the $x$ direction.
\vspace{0.1cm}%
From figure \ref{Vtot-qx-Barreiraq+q-}, the critical point is
defined as the inflexion point of the total potential energy and
mathematically it means,
\begin{eqnarray}
\frac{\partial V_{Tot}}{\partial \vec{q}}
\Bigg{\vert}_{\vec{q}_{c},\vec{r}_{c}}=0 \label{critPoint}
\end{eqnarray}
\begin{eqnarray}\label{critPoint1}
\Bigg{[}\frac{\partial^2 V_{Tot}}{\partial q_x^2}\frac{\partial^2
V_{Tot}}{\partial q_y^2} - \left(\frac{\partial^2 V_{Tot}}{\partial
q_y \partial q_x} \right)^2\Bigg{]}
\Bigg{\vert}_{\vec{q}_{c},\vec{r}_{c}}&=&0\\\label{critPoint2}
\frac{\partial^2 V_{Tot}}{\partial
q_z^2}\Bigg{\vert}_{\vec{q}_{c},\vec{r}_{c}}&>&0
\end{eqnarray}
The first condition, equation (\ref{critPoint}), is always satisfied
at equilibrium, it states that the total force on the particle must
vanish. The second condition, equation (\ref{critPoint1}), is
satisfied in a transition from a stable to an instable position on
the plane of the surface (critical points). It states that the
determinant of the Hessian matrix in the plane vanishes. It follows
from the fact that at these points, the slope of the force due to
the substrate equals the slope, $k_\alpha$, of the spring force. The
third condition, equation (\ref{critPoint2}), defines that the
particle is always in contact with the surface.

We can write the total potential as $V_{Tot}(\vec{q},\vec{r}) =
V_{Tot}(\vec{q}_{\parallel},\vec{r}_{\parallel},q_z,z)$, where
$\vec{q}_{\parallel}$ and $\vec{r}_{\parallel}$ are coordinates
parallel to the surface and the others correspond to the normal
components. We can expand the potential around the critical points,
$\vec{\chi}=(\vec{q}_{c},\vec{r}_{c})$, as
%
\begin{eqnarray}
V_{Tot} &\approx& A(\vec{r}_{\parallel},z) +
(\vec{r}_{\parallel}-\vec{r}_{\parallel c})\cdot \frac{\partial^{2}
V_{Tot}}{\partial \vec{r}_{\parallel}
\partial\vec{q}_{\parallel}} \Big{|}_{\vec{\chi}}
\cdot(\vec{q}_{\parallel}-\vec{q}_{\parallel c})
~~~~~~ \nonumber\\
 &+&\frac{1}{3!}\frac{\partial^3 V_{Tot}}{\partial
\vec{q}_{\parallel}^{3}} \Big{|}_{\vec{\chi}} \cdot
(\vec{q}_{\parallel}-\vec{q}_{\parallel c})^3 \nonumber \\
 &+&\frac{\partial^2 V_{Tot}}{\partial z \partial q_z}
\Big{|}_{\vec{\chi}} (z-z_c)(q_z-q_{zc})  \nonumber \\
 &+& 2\frac{\partial^{2} V_{Tot}}{\partial
\vec{q}_{\parallel}\partial q_z} \Big{|}_{\vec{\chi}} \cdot
(\vec{q}_{\parallel}-\vec{q}_{\parallel c})
(q_{z}-q_{zc})\nonumber\\
 &+& \frac{1}{2!}\frac{\partial^2
V_{Tot}}{\partial q_z^2}\Big{|}_{\vec{\chi}} (q_z-q_{zc})^2 +...
\label{TaylorSerie} ~~~~~~~
\end{eqnarray}
As a matter of clarity we write this potential in a more compact
form,
\begin{eqnarray}
V_{Tot} &\approx& A(x,y,z)+ C_1 (x-x_c)(q_x-q_{xc})\nonumber \\
    &+& C_2 (y-y_c)(q_y-q_{yc})+ C_3 (z-z_c)(q_z-q_{zc}) \nonumber \\
    &+& C_{13} (q_x-q_{xc})(q_z-q_{zc})+ (E_3/2)(q_z-q_{zc})^2 \nonumber \\
        &+& C_{23} (q_y-q_{yc})(q_z-q_{zc}) \nonumber \\
    &+& (E_1/6)(q_x-q_{xc})^3+ (E_2/6)(q_y-q_{yc})^3 + \dots ~~~~~
\label{expansion}
\end{eqnarray}
where the constants, $A$, $C_1$, $C_2$, $C_3$, $C_{13}$, $C_{23}$,
$E_1$, $E_2$ and $E_3$ will be explicitly written in the next
section, they are immediately obtained by direct comparison with
equation (\ref{TaylorSerie}). The main goal of this model is to show
that the dynamic energy barrier, $\Delta E$, which prevents the jump
of the tip from a stable equilibrium position on the surface to the
nearest neighboring one, is a function of the instantaneous lateral
force, $f_L$, and the instantaneous normal force, $f_z$. The energy
barrier $ \Delta E$ is defined as
\begin{eqnarray}
\Delta E = V_{Tot}(\vec{q}_{max},t) - V_{Tot}
(\vec{q}_{min},t)~~~~.\label{DefDeltaE}
\end{eqnarray}
The quantities $\vec{q}_{max}$ and $\vec{q}_{min}$ correspond to the
first minimum and maximum of the combined potential at time, $t$, it
is determined by the equilibrium condition $\partial V/\partial
\vec{q}=0$ \cite{Riedo}(See figure \ref{Vtot-qx-Barreiraq+q-}).

Using the equilibrium condition in the equation (\ref{expansion}) we
obtain,
\begin{eqnarray}
 \frac{E_1}{2}(q_x-q_{xc})^2 + C_{13}(q_z-q_{zc})+C_1 (x-x_c) & = & 0 \label{condx}\\
 \frac{E_2}{2}(q_y-q_{yc})^2 + C_{23}(q_z-q_{zc})+C_2 (y-y_c) & = & 0 \label{condy}\\
 E_3(q_z-q_{zc})+ C_{13}(q_x-q_{xc})+C_{23}(q_y-q_{yc})\nonumber \\
  +C_3 (z-z_c) & = & 0~~~~\label{condz}
\end{eqnarray}
substituting (\ref{condz}) in  (\ref{condx}), (\ref{condy}) and
holding only terms of first order in $C_{13}$ and $C_{23}$, we
obtain two quadratic equations in $q_x$, $q_y$ to solve, whose
solutions are
\begin{eqnarray}
 q_{x\pm}-q_{xc} = \pm\frac{1}{E_{1}^{1/2}}
  \sqrt{2C_1 (x_c^{'}-x)}  \label{SolQx}
\end{eqnarray}
\begin{eqnarray}
 q_{y\pm}-q_{yc} = \pm\frac{1}{E_{2}^{1/2}}
  \sqrt{2C_2 (y_c^{'}-y)}  \label{SolQy}
\end{eqnarray}
with
\begin{eqnarray}
C_1 x_c^{'}=C_1 x_c-(C_{13}/E_3) C_3(z_c-z)  &,&   \nonumber \\
C_2 y_c^{'}=C_2 y_c-(C_{23}/E_3) C_3(z_c-z)  &.&   \nonumber \\
\nonumber \label{Definitions}
\end{eqnarray}

Using the definition of $\Delta E$, and equations from
(\ref{expansion}) to (\ref{SolQy}) we can get the energy barrier as
\begin{eqnarray}
\Delta E=\frac{2^{\frac{7}{2}}}{3} \Bigg{\{} \frac{[C_1
(x_c^{'}-x)]^{3/2}}{E_{1}^{1/2}}+ \frac{[C_2
(y_c^{'}-y)]^{3/2}}{E_{2}^{1/2}} \Bigg{\}}.
\label{AnaEnerBarXY}
\end{eqnarray}
Defining,
\begin{eqnarray}
f_x(t)=C_1 x(t) &;& F_{xc}=C_1 x_c ~;~ F_{xc}^{'}=C_1 x_c^{'}~~, \nonumber \\
f_y(t)=C_2 y(t) &;& F_{yc}=C_2 y_c ~;~ F_{yc}^{'}=C_2 y_c^{'}~~, \nonumber \\
f_z(t)=C_3 z(t) &;& F_{zc}=C_3 z_c~~~~,  \nonumber
\label{Definitions1}
\end{eqnarray}
\begin{eqnarray}
f_x(t)=f_L(t)\cos(\theta) &;&  f_y(t)=f_L(t)\sin(\theta)~~, \nonumber \\
F_{xc}=F_{lc}\cos(\theta) &;&  F_{yc}=F_{lc}\sin(\theta)~~, \nonumber \\
F_{lc}=\sqrt{F_{xc}^2+F_{yc}^2} &;& f_L(t)=\sqrt{f_x(t)^2+f_y(t)^2}~~, \nonumber\\
F_{xc}^{'}=F_{lc}^{'}\cos(\theta) &;&
F_{yc}^{'}=F_{lc}^{'}\sin(\theta)~~, \nonumber \label{Definitions11}
\end{eqnarray}
we find an energy barrier which is similar to that found in other
works \cite{Sang,PerssonWear,ASocoliuc},
\begin{eqnarray}
\Delta E(t)=\lambda_1(\theta) \left [ F_{lc}^{'}(f_z(t))-f_L(t)
\right ]^{3/2} ~~, \label{AnaEnerBar}
\end{eqnarray}
where,
%
\begin{eqnarray}
\mu_{0x}&=&\frac{C_{13}}{E_3} ~~;~~ \mu_{0y}=\frac{C_{23}}{E_3}~~,  \label{Definitions2.2} \\
\lambda_1(\theta)&=&\frac{2^{\frac{7}{2}}}{3} \left\lbrace
\frac{|\cos(\theta)|^{\frac{3}{2}}}{E_{1}^{\frac{1}{2}}}+
\frac{|\sin(\theta)|^{\frac{3}{2}}}{E_{2}^{\frac{1}{2}}}
\right\rbrace~~,  \label{Definitions2.3}
\end{eqnarray}
\begin{widetext}
\begin{eqnarray}
F_{lc}^{'}(f_z(t))&=&\sqrt{F_{lc}^2+(\mu_{0x}^2+
\mu_{0y}^2)(f_z(t)-F_{zc})^2+2F_{lc}(\mu_{0x}\cos(\theta)+\mu_{0y}\sin(\theta))(f_z(t)-F_{zc})}
\label{Definitions2.1}
\end{eqnarray}
\end{widetext}
We note that $\Delta E$ presents a dependence with the instantaneous
normal force, $f_z$, and the angle $0\leq\theta\leq\pi$ between the
direction $x$ and the force pushing the cantilever. Now we will
study the consequences of this dependence in the friction force. At
zero temperature, the lateral force required to the tip jumping from
a minimum to another is $f_L(t)=F_{lc}^{'}(f_z(t))$ (i.e., the force
corresponding to $ \Delta E=0$). At finite temperature, $T$, the
occurrence of thermally activated transitions between two minima,
when $ \Delta E > 0$, leads to $f_L(t)<F_{lc}^{'}(f_z(t))$ and the
probability that the tip does not jump, $p(t)$, is described by the
master equation \cite{Gnecco,EGneccoWear254,ASocoliuc,Riedo}:
\begin{equation}
\frac{\partial p(t)}{\partial t}=-\nu_{0} \exp\left[-\beta \Delta
E(t)\right]p(t) \label{SimpMasEq}
\end{equation}
where $\beta=1/k_b T$, $\nu_{0}$ is the jump frequency transition
and $k_b$ is the Boltzmann constant. Using the condition for the
maximum jumping probability, $d^2 p/dt^2=0$, the following
expression has to be satisfied:
\begin{equation}
\Delta E=-\frac{1}{\beta}\ln\left(-\frac{\beta}{\nu_0}\frac{d \Delta
E}{dt}\right) \label{Condicao}
\end{equation}
Then
\begin{eqnarray}\label{Df_ldt}
\frac{d f_L}{dt}&=&\left[\frac{\partial f_L}{\partial
f_x}\right]\left[\frac{\partial f_x}{\partial
x}\right]\left[\frac{\partial x}{\partial t}\right]+
\left[\frac{\partial f_L}{\partial f_y}\right]\left[\frac{\partial
f_y}{\partial y}\right] \left[\frac{\partial y}{\partial t}\right]\nonumber\\
&=&\left[\frac{f_x}{f_L}\right][\kappa_{eff}^{x}][v_x]+
\left[\frac{f_y}{f_L}\right][\kappa_{eff}^{y}][v_y]\nonumber \\
&=&(\kappa_{eff}^{x}\cos(\theta)^2+\kappa_{eff}^{y}\sin(\theta)^2)v_l
=\varphi v_l \nonumber
\end{eqnarray}
\begin{eqnarray}\label{DF_lcdt}
\frac{d F_{lc}^{'}}{dt}&=&\left[\frac{\partial F_{lc}^{'}}{\partial
f_z}\right]\left[\frac{\partial f_z}{\partial
z}\right]\left[\frac{\partial z}{\partial t}\right]
=\left[\frac{\partial F_{lc}^{'}}{\partial
f_z}\right]\kappa_{eff}^{z}v_z \nonumber\\
&=&\gamma\kappa_{eff}^{z}v_z\nonumber
\end{eqnarray}
\begin{eqnarray}\label{DeltaEdt}
\frac{d \Delta E}{dt}%
 &=&-\frac{3}{2}\lambda_1 \left [ F_{lc}^{'}(f_z)-f_L \right ]^{1/2}
\left(\frac{d f_L}{dt}-\frac{d F_{lc}^{'}}{dt}\right)\nonumber\\
 &=&-\frac{3}{2}\lambda_1 \left [ F_{lc}^{'}(f_z)-f_L \right ]^{1/2}
|\varphi v_{l} +\gamma \kappa_{eff}^{z} v_z|~~~~
\end{eqnarray}
where,
\begin{eqnarray}
\varphi(\theta,f_z)=\kappa_{eff}^{x}\cos(\theta)^2+\kappa_{eff}^{y}\sin(\theta)^2 ~~~~\label{Phi}\\
v_{x}=v_{l}\cos(\theta) ~~;~~ v_{y}=v_{l}\sin(\theta)~~;~~
v_{l}=\sqrt{v_{x}^2+v_{y}^2}\nonumber
\end{eqnarray}
\begin{eqnarray}
\gamma(\theta,f_z)=\frac{1}{F_{lc}^{'}}\Big[(\mu_{0x}^2+ \mu_{0y}^2)
(f_z(t)-F_z^{*})\nonumber\\
+F_{lc}(\mu_{0x}\cos(\theta)+\mu_{0y}\sin(\theta))\Big]
\label{Gamma},
\end{eqnarray}
and $\kappa_{eff}^{\alpha}$ is the effective stiffness of contact
\cite{lantz:970} in the directions $\alpha=x,y,z$. It is important
to stand out that $\kappa_{eff}^{x,y}$ is a function of the normal
force. This dependence, will appear in a crucial way in the
deduction of the theoretical coefficient of friction that we will
present. The effective stiffness of contact can be experimentally
determined by measuring the inclination of the ``stick region'',
which is the region were the tip is in a stable position before the
equation (\ref{critPoint1}) is satisfied in the friction force loop
\cite{carpick:1548}. Thus, the effective stiffness of contact is
defined as,
\begin{eqnarray}\label{EffStiffnessX1}
\kappa_{eff}^{x}=\frac{\partial f_x}{\partial x}=\left( \frac{1}{
\kappa_{contact}^{x}} + \frac{1}{k_x} \right)^{-1},
\end{eqnarray}
where $\kappa_{contact}^{x}$ is the stiffness of contact which is a
function of the normal force \cite{PhysRevB.52.14976,FogdenWhite}
and $k_x$(equation (\ref{SpringTensor})), is the spring constant of
the cantilever. Using equations (\ref{AnaEnerBar}), (\ref{Condicao})
and (\ref{DeltaEdt}) we obtain the lateral force,
\begin{widetext}
\begin{eqnarray}
f_L&=& F_{lc}^{'}(f_z) -\Bigg{\{} \frac{k_bT}{\lambda_1(\theta)}
\Bigg{[}\ln\left( \frac{v_0(f_z)}{|\varphi v_{l} -\kappa_{eff}^{z}
\gamma v_z|}
\right)  
-
\frac{1}{2}\ln\Bigg{(}1-\frac{f_L}{F_{lc}^{'}(f_z)}\Bigg{)}\Bigg{]}\Bigg{\}}^{2/3}
\label{latforce}
\end{eqnarray}
\end{widetext}
where
\begin{eqnarray}
v_0(f_z)=\frac{2\nu_0k_b T }{3\lambda_1(\theta)(F_{lc}^{'})^{1/2}}
\label{vzero},
\end{eqnarray}

This result has a similar form to those presented by Riedo et. al
\cite{Riedo} with three interesting physical differences: 1) the
explicitly normal force dependence on $F_{lc}^{'}(f_z)$ and
$v_0(f_z)$ as indicated by experimental results \cite{Riedo}; 2)the
angular dependence of the friction force relative to the periodic
crystalline structure of the sliding surface; 3) the logarithmic
normal velocity dependence of the friction force, as Jeon et. al.
\cite{SangminJeon} have shown experimentally and numerically
simulated using a single particle model. Note that by introducing a
normal force dependence in the friction force it is natural to think
of the coefficient of friction as defined by
\begin{widetext}
\begin{eqnarray}
\mu(t)=\frac{\partial f_L(t) }{\partial f_z(t)}=
\gamma+2\left(\frac{k_b T}{\lambda_1}\right)^{2/3} \times
\frac{\left[\frac{\partial\varphi}{\partial f_z} v_{l}
-\kappa_{eff}^{z} \frac{\partial\gamma}{\partial f_z}
v_z\right]}{[\varphi v_{l} -\kappa_{eff}^{z} \gamma v_z]}
\frac{\Bigg{\{}\ln\left( \frac{v_0}{|\varphi v_{l} -\kappa_{eff}^{z}
\gamma v_z|} \right) - \frac{1}{2} \ln \Bigg{(}1 -
\frac{f_L}{F_{lc}^{'}} \Bigg{)}\Bigg{\}}^{2/3} }{\Bigg{\{}1+3\left[
\ln\left( \frac{v_0}{|\varphi v_{l} -\kappa_{eff}^{z} \gamma v_z|}
\right) - \frac{1}{2} \ln\Bigg{(}1-\frac{f_L}{F_{lc}^{'}}
\Bigg{)}\right] \Bigg{\}}} \label{CeffFric}
\end{eqnarray}
\end{widetext}
where
\begin{eqnarray}
\frac{\partial\varphi}{\partial f_z}=\frac{\partial \kappa_{eff}^{x}
}{ \partial f_z} \cos(\theta)+\frac{\partial \kappa_{eff}^{y} }{
\partial f_z} \sin(\theta) \label{DelphiDelfz}
\end{eqnarray}
\begin{eqnarray}
\frac{\partial\gamma}{\partial f_z}=\frac{ F_{lc}^{2}
\left[\mu_{0x}\sin(\theta)-\mu_{0y}\cos(\theta)\right]^2 }{
(F_{lc}^{'})^{3} } \label{DelgammaDelfz}
\end{eqnarray}
Observe that this equation gives a friction coefficient that is a
function of velocity, temperature, normal force and the direction
that the cantilever is dragged in the surface. In the next sections
we show this results for a selected potential and we compare it with
MD simulations. It is important to note that if we know the
potential between the surface and the tip atoms we can predict the
value of the friction coefficient with no adjustable parameters.

In the next section we will use a total potential to calculate the
values of the parameters defined in the model and show that the
theoretical friction force agree very well with experimental results
\cite{Riedo,EuiSungYoon}.
%
%
%
%
\section{Surface-tip's Potential and Parameters of the Analytic Model\label{SurfPotParaAnaliModel}}
The model we developed has no adjustable parameters. Once the
potential energy is given, we are able to calculate the friction
force. For a well behaved surface, Steele\cite{Steele} has derived a
model potential which is very fair in describing the interaction
with a periodic surface
\begin{eqnarray} \label{Uint}
V_{int}(\vec{q}_{\parallel},q_z) = V_{0}(q_z) +
V_{1}(q_z)\sum_{\vec{G}}\cos(\vec{G}\cdot\vec{q}_{\parallel}),
\end{eqnarray}
where $\vec{q}_{\parallel}=(q_x,q_y)$ are the coordinates of the
tip's atom parallel to the substrate and $\vec{G}$ is the set of the
six shortest reciprocal lattice vectors of the substrate. The first
term in equation (\ref{Uint}) describes the mean interaction of the
atoms with the substrate, and the second term describes the periodic
corrugation potential. Expressions for $V_0$ and $V_1$ were derived
by Steele\cite{Steele}, assuming that the substrate potential
$V_{int}(\vec{q})$ is a sum of LJ potentials between one film atom
and all of the atoms in the substrate. The parameters of the LJ
potential can nowadays be experimentally determinated by DFS
\cite{PRL93} experiments for a specific material. Other expressions
for $V_0$ and $V_1$, where described by some authors such as Persson
et. al. \cite{PerssonNitzan}, Liebsch et. al. \cite{Liebsch} that
consider these terms composed by exponentials and Tomassone et. al.
\cite{Tomassone} consider expressions that give a correct
description of the interaction of a metallic surface with a noble
gas atom. In this work we use the expression derived by Steele. At
this point we need to define the surface arrangement of particles
which we are interested to work with. We use the regular BCC(001)
surface, so that the total potential energy (Equation
(\ref{PotUTot})) becomes
\begin{eqnarray} \label{Uint1}
V_{Tot}= \frac{1}{2}\left[ (\vec{q}-\vec{r}) \cdot
\overleftrightarrow{k} \cdot (\vec{q}-\vec{r}) \right] +V_{0}(q_z)\nonumber \\
+2V_{1}(q_z)[\cos(G_x q_x)+\cos(G_y q_y)]~~.
\end{eqnarray}
Using the conditions imposed by the equations (\ref{critPoint}) to
(\ref{critPoint2}) in equation (\ref{Uint1})  one obtain,
\begin{eqnarray}
k_x(q_{xc}-x_c) -2V_{1}(q_{zc})\sin(G_x q_{xc})G_x =0~~,\label{Del1Uint1} \\
k_y(q_{yc}-y_c) -2V_{1}(q_{zc})\sin(G_y q_{yc})G_y =0~~,\label{Del1Uint1-1} %
\end{eqnarray}
\begin{eqnarray}
k_z(q_{zc}-z_c)+ V^{'}_{0}(q_z) \nonumber \\
+2V^{'}_{1}(q_{zc})[\cos(G_x q_{xc})+\cos(G_y q_{yc})]=0\label{Del1Uint11-1}~~,%
\end{eqnarray}
\begin{eqnarray} \label{Del2Uint1}
\frac{\partial^2 V_{Tot}}{\partial q_{x}^{2}}&=& k_x \nonumber
-2V_{1}(q_z)\cos(G_x q_x)G_{x}^{2} ~~,\\\nonumber \frac{\partial^2
V_{Tot}}{\partial q_{y}^{2}}&=& k_y -2V_{1}(q_z)\cos(G_y
q_y)G_{y}^{2} ~~,\\\nonumber \frac{\partial^2 U_{Tot}}{\partial
q_{x}
\partial q_{y}}&=& 0~~,\nonumber
\end{eqnarray}
\begin{eqnarray}\nonumber
\Bigg{[}\frac{\partial^2 V_{Tot}}{\partial q_x^2}\frac{\partial^2
V_{Tot}}{\partial q_y^2} - (\frac{\partial^2 V_{Tot}}{\partial q_y
\partial q_x})^2\Bigg{]}\Bigg{\vert}_{\vec{q}_{c},\vec{r}_{c}}&=&0\\\nonumber
\end{eqnarray}
\begin{eqnarray}
\left[ k_x-2V_{1}(q_{zc})\cos(G_x q_{xc})G_{x}^{2}\right] \times
\nonumber\\ \left[ k_y-2V_{1}(q_{zc})\cos(G_y
q_{yc})G_{y}^{2}\right]  &=& 0\label{UsingCritPoint1}
\end{eqnarray}
Solving these equations we get,
\begin{eqnarray}\nonumber
q_{xc}=\frac{1}{G_x}\arccos(\eta_x) ~~;~~ \sin(G_x
q_{xc})=\pm\sqrt{1-\eta_x^2}~~,\\
q_{yc}=\frac{1}{G_y} \arccos(\eta_y)  ~~;~~ \sin(G_y
q_{yc})=\pm\sqrt{1-\eta_x^2}~~, \label{UsingCritPoint2}
\end{eqnarray}
where
\begin{eqnarray}
\label{Eta_xy}
\eta_x=\frac{k_x}{2V_{1}(q_{zc})G_{x}^{2}}~~,~~~~~~~~~~~~
\eta_y=\frac{k_y}{2V_{1}(q_{zc})G_{y}^{2}}~~.
\end{eqnarray}
As we can see, equations (\ref{UsingCritPoint2}) and (\ref{Eta_xy})
define a transition from a stable to instable tip's position, or in
other words a transition from the stick to the slip state. Using
this information we are able to explicitly calculate the parameters
defining the friction force of the analytical model presented above.
With this in mind, we explicitly define and calculate the parameter
$C_1$, $C_2$, $C_3$, $C_{13}$, $C_{23}$, $E_1$, $E_2$ and $E_3$
which led to equation (\ref{expansion}). From equation (\ref{Uint1})
we obtain,
\begin{eqnarray}\label{DefContC1C2C3}
C_1 &=& \frac{\partial^{2} V_{Tot}}{\partial q_x \partial x}
\Bigg{\vert}_{\vec{q}_{c},\vec{r}_{c}} = -k_x \nonumber \\%
C_2 &=& \frac{\partial^{2} V_{Tot}}{\partial q_x \partial x}
\Bigg{\vert}_{\vec{q}_{c},\vec{r}_{c}} = -k_y \nonumber \\%
C_3 &=& \frac{\partial^{2} V_{Tot}}{\partial q_x \partial x}
\Bigg{\vert}_{\vec{q}_{c},\vec{r}_{c}} =
-k_z.\nonumber %
\end{eqnarray}

Using equations (\ref{Del1Uint1}), (\ref{Del1Uint1-1}) ,
(\ref{Del1Uint11-1}), (\ref{UsingCritPoint2}) and (\ref{Eta_xy}) we
get
\begin{eqnarray}\label{FLC-FZC}
F_{xc}&=&C_{1} x_c = -\frac{k_x}{G_x}\left(\arccos(\eta_x)+\frac{1}{\eta_x}\sqrt{1-\eta_x^2}\right) \\
F_{yc}&=&C_{2} y_c = -\frac{k_y}{G_y}\left(\arccos(\eta_y)+\frac{1}{\eta_y}\sqrt{1-\eta_y^2}\right) \\
F_{lc}&=&\sqrt{F_{xc}^2+F_{yc}^2}    \\
F_{zc}&=&C_3 z_c=-k_z q_{zc}-V^{'}_{0}-\frac{V^{'}_{1}}{V_{1}}
\left[ \frac{k_x}{G_{x}^{2}}+\frac{k_y}{G_{y}^{2}}\right]~~~~
\end{eqnarray}

\begin{figure}[htbp]
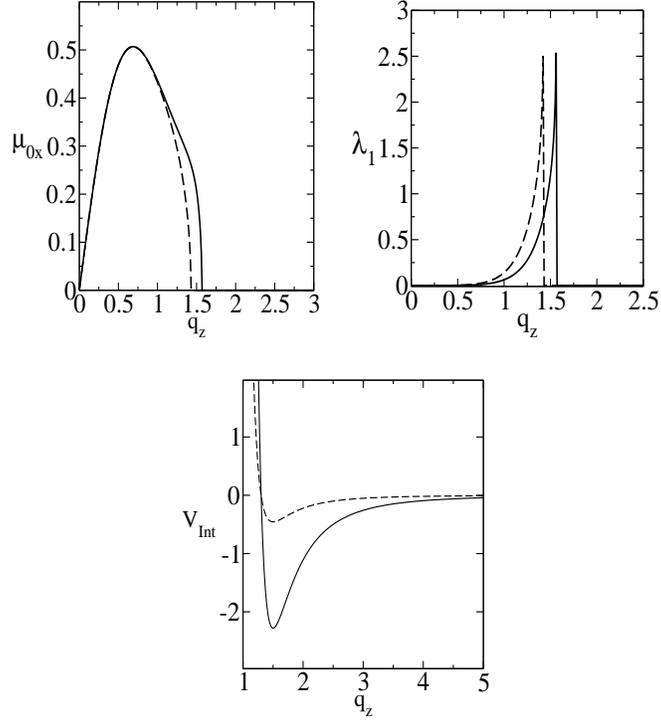

\begin{center}
\includegraphics[width=4.10cm,height=4.50cm]{figs/Fig3a.eps}\hspace* {0.3cm}
\includegraphics[width=4.10cm,height=4.50cm]{figs/Fig3b.eps}\vspace{0.3cm} \\\vspace{0.2cm}
\includegraphics[width=4.10cm,height=4.50cm]{figs/Fig3c.eps}
\end{center}\vspace* {0.0cm}
\caption{The parameters $\mu_{0x}=\mu_{0}$, $\lambda_1$ and
$V_{int}(\vec{q}_{\parallel}=(a_1,a_1),q_z)$ plotted as a function
of $q_{z}$. Here $\theta=0$, $\sigma_{ts}=1.2\sigma_{ss}$,
$\varepsilon_{ts}=0.5\varepsilon_{ss}$ (full),
$0.1\varepsilon_{ss}$(dashed). } \label{mi0-lambda1-Uint-qzc}
\end{figure}

To calculate $\mu_{0}$ and $\lambda_1(\theta)$ we start by
calculating $C_{13}$, $C_{23}$ and $E_3$. From equations
(\ref{Uint1}), (\ref{UsingCritPoint2}) and (\ref{Eta_xy})  we can
write
\begin{eqnarray} \label{DefContC13C23E3}
C_{13}&=& \frac{\partial^{2} V_{Tot}}{\partial q_x \partial q_z}
\Bigg{\vert}_{\vec{q}_{c},\vec{r}_{c}} = -2U^{'}_{1}(q_{zc})G_x\sqrt{1-\eta_{x}^{2}} \\
C_{23}&=& \frac{\partial^{2} V_{Tot}}{\partial q_y \partial q_z}
\Bigg{\vert}_{\vec{q}_{c},\vec{r}_{c}} = -2U^{'}_{1}(q_{zc})G_y\sqrt{1-\eta_{y}^{2}} \\
E_3 &=& \frac{\partial^{2} V_{Tot}}{\partial q_{z}^{2}}
\Bigg{\vert}_{\vec{q}_{c},\vec{r}_{c}} \nonumber\\
&=& k_z + V^{''}_{0}+\frac{V^{''}_{1}}{V_{1}} \left[
\frac{k_x}{G_{x}^{2}}+\frac{k_y}{G_{y}^{2}}\right] \neq 0~~~~
\end{eqnarray}
Using these results in equations (\ref{Definitions2.2}) we obtain,
\begin{eqnarray} \label{mu0xy}
\mu_{0x}&=&\frac{-2V^{'}_{1}(q_{zc})G_x\sqrt{1-\eta_{x}^{2}}}{|E_3|},\\
\mu_{0y}&=&\frac{-2V^{'}_{1}(q_{zc})G_y\sqrt{1-\eta_{y}^{2}}}{|E_3|}.
\end{eqnarray}
From equations (\ref{Uint1}), (\ref{UsingCritPoint2}) and
(\ref{Eta_xy}) we obtain $E_1$ and $E_2$ by,
\begin{eqnarray} \label{DefContE1E2}
E_1&=&\frac{\partial^3 V_{Tot}}{\partial q_{x}^{3}}
\Bigg{\vert}_{\vec{q}_{c},\vec{r}_{c}} =
\frac{k_x G_x}{\eta_x}\sqrt{1-\eta_x^2} \\
E_2&=&\frac{\partial^3 V_{Tot}}{\partial q_{y}^{3}}
\Bigg{\vert}_{\vec{q}_{c},\vec{r}_{c}} = \frac{k_y
G_y}{\eta_y}\sqrt{1-\eta_y^2}
\end{eqnarray}
where as defined by equation (\ref{Definitions2.3}) we can calculate
$\lambda_1(\theta)$ as,
\begin{eqnarray} \label{DefContE1E23}
\lambda_1(\theta)=\frac{2^{\frac{7}{2}}}{3} \left\lbrace
\left(\frac{\eta_{x}}{k_x G_x}\right)^{1/2}
\frac{|\cos(\theta)|^{\frac{3}{2}}}{(1-\eta_x^2)^{1/4}}+\right.\nonumber\\
\left. \left(\frac{\eta_{y}}{k_y G_y}\right)^{1/2}
\frac{|\sin(\theta)|^{\frac{3}{2}}}{(1-\eta_y^2)^{1/4}}
\right\rbrace~~.
\end{eqnarray}
Note that all the parameters $F_{lc}$, $F_z^{*}$, $\mu_{0}$, and
$\lambda_1$ are functions of the critical point $q_{zc}$. Using the
functions $V_{0}(q_z)$ and $V_{1}(q_z)$ of the Steele potential
defined by \cite{Steele},
\begin{eqnarray} \label{U0Stelle}
V_{0}&=&\frac{2\pi q \sigma_{ts}^6 \varepsilon_{ts}}{a_{s}}
\sum^{\infty}_{p=0}\frac{1}{(q_z+p\Delta q_z)^4}\left( \frac{2
\sigma_{ts}^6}{5(q_z+p\Delta q_z)^{6}}-1 \right)\nonumber\\
\end{eqnarray}
\begin{eqnarray} \label{U0lStelle}
V_{0}^{'}&=&-\frac{8\pi q \sigma_{ts}^6 \varepsilon_{ts}}{a_{s}}
\sum^{\infty}_{p=0}\frac{1}{(q_z+p\Delta
q_z)^5}\left( \frac{ \sigma_{ts}^6}{(q_z+p\Delta q_z)^{6}}-1 \right)\nonumber\\
\end{eqnarray}
\begin{eqnarray} \label{U0llStelle}
V_{0}^{''}&=&\frac{8\pi q \sigma_{ts}^6 \varepsilon_{ts}}{a_{s}}
\sum^{\infty}_{p=0}\frac{1}{(q_z+p\Delta q_z)^6}\left( \frac{11
\sigma_{ts}^6}{(q_z+p\Delta q_z)^{6}}-5
\right)\nonumber\\
\end{eqnarray}
where $p$ is an integer, $q$ is the total number of atoms per unit
surface cell, $\Delta q_z$ is the distance between planes,
$a_{s}=a_1^2$, is the area of the unit lattice cell,
$a_{1}=\sigma_{ss}$, $\sigma_{ss}$ is the nearest neighbor in the
solid and $\sigma_{ts}$ and $\varepsilon_{ts}$ are the parameters of
a $(6-12)$ Lennard-Jones potential between the tip and the surface
atoms, and

\begin{eqnarray} \label{U1Stelle}
V_{1}(q_z)=\frac{2\pi \sigma_{ts}^6 \varepsilon_{ts}}{a_{s}} \left[
\frac{\sigma_{ts}^6}{30} \left( \frac{g_{1}}{2q_z} \right)^5
K_{5}(g_{1}q_z) \right. \nonumber\\
\left. -2\left(\frac{g_{1}}{2q_z}\right)^2 K_{2}(g_{1}q_z) \right]
\end{eqnarray}
\begin{eqnarray} \label{U1lStelle}
V_{1}^{'}(q_z)=\frac{-2\pi \sigma_{ts}^6 \varepsilon_{ts}}{a_{s}}
\left\{ \frac{\sigma_{ts}^6}{30}\left( \frac{g_{1}}{2q_z} \right)^5
\times \right.\nonumber\\ \left. \times\left[\frac{10
K_{5}(g_{1}q_z)}{q_z}+g_1K_{4}(g_{1}q_z)
\right]\right.\nonumber\\
-\left. 2\left(\frac{g_{1}}{2q_z}\right)^2
\left[\frac{4K_{2}(g_{1}q_z)}{q_z}+g_1K_{1}(g_{1}q_z)\right]
\right\}
\end{eqnarray}
\begin{eqnarray} \label{U1llStelle}
V_{1}^{''}(q_z)=\frac{2\pi \sigma_{ts}^6 \varepsilon_{ts}}{a_{s}}
\left\{ \frac{\sigma_{ts}^6}{30}\right. \left( \frac{g_{1}}{2q_z}
\right)^5 \left[\frac{110 K_{5}(g_{1}q_z)}{q_z^2}\right. \nonumber\\
\left.
+\frac{19 g_1 K_{4}(g_{1}q_z)}{q_z} + g_{1}^{2} K_{3}(g_{1}q_z) \right] \nonumber\\
- \left. 2\left(\frac{g_{1}}{2q_z}\right)^2 \left[\frac{20
K_{2}(g_{1}q_z)}{q_z^2}+\frac{7 g_1 K_{1}(g_{1}q_z)}{q_z} \right.\right.\nonumber\\
+ g_{1}^{2} K_{0}(g_{1}q_z) \Big ] \Big \}
\end{eqnarray}
where $K_{n}$ is the modified Bessel function of second kind and
$g_1=G_x=G_y=2\pi/a_1$. Using the equations from (\ref{U0Stelle}) to
(\ref{U1llStelle}) we plot in figure \ref{mi0-lambda1-Uint-qzc} the
parameters $\mu_{0x}=\mu_{0}$, and $\lambda_1$ and
$V_{int}(\vec{q}_{\parallel}=(a_1,a_1),q_z)$ as a function of
$q_{z}$, for values of $\eta_x=\eta_y<1$ since we are interested in
real values of the parameters.

\vspace* {0.5cm}

\subsection{Theoretical prediction of the friction force and coefficient of friction}
\begin{figure}[htbp]
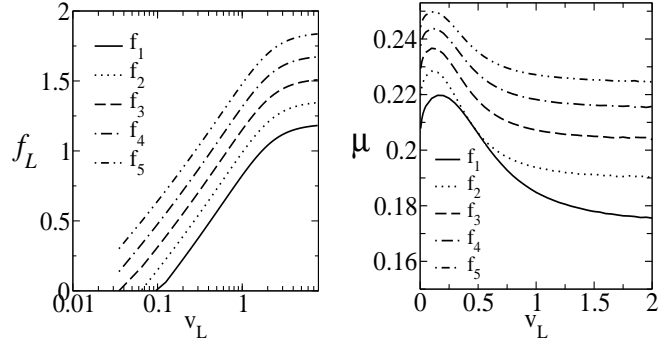

\begin{center}
\includegraphics[width=4.1cm,height=4.50cm]{figs/Fig4a.eps}\hspace* {0.30cm}
\includegraphics[width=4.1cm,height=4.50cm]{figs/Fig4b.eps}
\end{center}\vspace* {0.0cm}
\caption{Theoretical friction force as a function of the velocity
for different values of the normal force given by the equation
(\ref{latforce}). Here from bottom to top are shown $f_{N1}<$
$f_{N2}<$ $f_{N3}<$ $f_{N4}<$ $f_{N5}$ respectively.}
\label{Plot-FLTeo-v}
\end{figure}

In the figure \ref{mi0-lambda1-Uint-qzc} we show the parameters
$\mu_{0x}$, $\lambda_1$ and $V_{int}$ calculated by using the
equations (\ref{mu0xy}), (\ref{DefContE1E23}) and (\ref{Uint})
respectively. Using those results we can calculate the friction
force, ($f_L$), and the coefficient of friction, ($\mu$). The figure
\ref{Plot-FLTeo-v} shows ($f_L$) and ($\mu$) as a function of the
relative velocity $v_L$ for several values of the of the normal
force, ($f_z$). We note that these results agree quite well with the
experimental data of {\it Riedo et.al} \cite{Riedo}.
%
\vspace* {0.50cm}
%
%
\vspace* {1.0cm}
\begin{figure}[htbp]
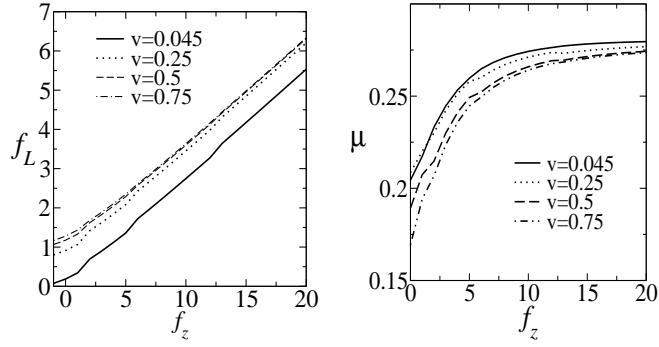

\begin{center}
\includegraphics[width=4.1cm,height=4.50cm]{figs/Fig5a.eps}\hspace* {0.30cm}
\includegraphics[width=4.1cm,height=4.50cm]{figs/Fig5b.eps}
\end{center}\vspace* {0.0cm}
\caption{Friction force(left) and friction coefficient(right) as a
function of the normal force for different values of the velocity.
Here from bottom to top are shown $v_{L}=0.045$, $v_{L}=0.25$,
$v_{L}=0.5$, $v_{L}=0.75$. } \label{Plot-FLTeo-Fz}
\end{figure}
%

In the figure \ref{Plot-FLTeo-Fz} we present the dependence of the
friction force (left) and the coefficient of friction (right) with
the normal force $(f_z)$. We note that the dependence of the
coefficient of friction (right) with the normal force $(f_z)$ agree
quite well with the experimental data of {\it Eui-Sung Yoon et.al}
\cite{EuiSungYoon}.

\begin{figure}[htbp]
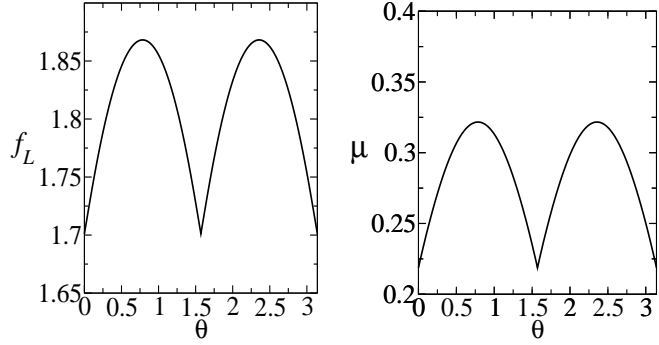

\begin{center}
\includegraphics[width=4.1cm,height=4.50cm]{figs/Fig6a.eps}\hspace* {0.30cm}
\includegraphics[width=4.1cm,height=4.50cm]{figs/Fig6b.eps}
\end{center}\vspace* {0.0cm}
\caption{Dependence of the friction force(left) and the friction
coefficient(right) on the direction of the sweepings, $\theta$,
relative to the (100) direction on the surface.}
\label{Fl-mi-theta-Teo}
\end{figure}

In the figure \ref{Fl-mi-theta-Teo} we present the dependence of the
friction force (left) and the coefficient of friction (right) with
the pushing angle, $\theta$, relative to the direction (100) of the
lattice. It is important to note that this result is a particular
case for the potential defined by the equation (\ref{Uint1}). It is
expected that the effect strongly depends on the symmetries of the
surface.
%
\vspace* {0.50cm}
%
%
%
%
\section{Simulation \label{SimDetails}}
In this section we present our simulation for the system discussed
above. The simulation is carried out using molecular dynamics (MD).
The simulational arrangement is as shown in figure \ref{FFMesqSim},
which represents a tip interacting with a surface. The tip is
simulated by a single particle. Three springs are attached to the
particle, two parallels to the surface plan ($x,y$ directions) and
the other perpendicular to the surface plan ($z$ direction). This
arrangement allow us to measure the forces, normal ($f_z$) and
parallel to the surface ($f_x,f_y$). The surface is represented by
an arrangement of particles which interact with each other through a
Lennard-Jones (LJ) $(6-12)$ potential
\begin{widetext}
\begin{equation}
\Phi_{i,j}(r_{i,j}) = \left \{
\begin{array}{lcc}
  \phi_{i,j}(r_{i,j}) - \phi_{i,j}(r_{c}) - (r_{i,j} - r_{c})
  \left( \frac{d\phi_{i,j}(r_{i,j})}{dr_{i,j})}
  \right)_{r_{i,j}=r_c} & if & r_{i,j} < r_{c} \\
  0 & if & r_{i,j} > r_{c} \\
\end{array} \right.
\label{equation1}
\end{equation}
\end{widetext}
\begin{equation}
\phi_{i,j}(r_{i,j}) = \epsilon_{i,j} \left[ \left(
\frac{\sigma_{i,j}}{r_{i,j}} \right)^{12} -\left(
\frac{\sigma_{i,j}}{r_{i,j}} \right)^{6} \right]
\end{equation}
The indexes $i$ and $j$ stands for particles at position
$\overrightarrow{r_{i}}$ and $\overrightarrow{r_{j}}$ respectively,
and $1 \leq i,j \leq N$, where $N$ is the total number of particles,
$\sigma_{i,j}=\sigma_{ss}$, $\epsilon_{i,j}=\epsilon_{ss}$ between
surface atoms and $\sigma_{i,j}=\sigma^{'}_{ts}\sigma_{ss}$,
$\epsilon_{i,j}=\epsilon^{'}_{ts}\epsilon_{ss}$ between the tip and
surface atoms.
A cutoff is introduced in the potential in order to accelerate the
simulation. If the force on a particle is found by summing
contributions from all particles acting upon it, then this
truncation limits the computation time to an amount proportional to
the total number of particles $N$. Of course, this truncation
introduces discontinuities both in the potential and the force. To
smooth these discontinuities we introduce the constant term
$\phi(r_c)$. Another term $\left( {d\phi_{i,j}(r_{i,j})}/{dr_{i,j}}
\right)_{r_{i,j}=r_c}$ is introduced to remove the force
discontinuity. Particles in the simulation move according Newton's
law of motion, which generates a set of $3N$ coupled equations of
motion  which are solved by increasing forward in time the physical
state of the system in small time steps of size $\delta
t=10^{-3}\sigma_{ss}\sqrt{(m/\epsilon_{ss})}$. The resulting
equations are solved by using Beeman's method of integration. In
order to improve the simulations we use a Verlet table and cell
division method. The temperature, $T$, of the surface can be
controlled by using a velocity renormalization scheme
\cite{Allen,Beeman,Berendsen,pablo1,pablo2,flavio1,flavio2,Rapaport}.
From the equipartition theorem we can write that
\begin{equation}
\langle v^2 \rangle = 3\frac{k_B}{m}T. \label{equation2}
\end{equation}
We want to control the value of $\langle v^2 \rangle$ to correspond
to a chosen temperature $T_f$. By initializing the system with
$\langle v^2 \rangle_0$ we multiply each velocity by a factor
$\alpha_0$
\begin{equation}
\alpha_0 = \sqrt{\frac{m}{3k_B}\frac{{\langle v^2 \rangle}_0}{T_f}}.
\label{equation3}
\end{equation}
By evolving in time the system we can create a sequence $\{\alpha_n
\}$, such that after a finite number of time steps the temperature
of the system converges to $T_f$. We measure the time $t$, and
temperature $T$, in units of $\sigma_{ss} \sqrt{m/\epsilon_{ss}}$
and $\epsilon_{ss}/k_B$ respectively.

%
\subsection{Numerical background}
\begin{figure}[htbp]
  \includegraphics[width=6.10cm,height=5.10cm]{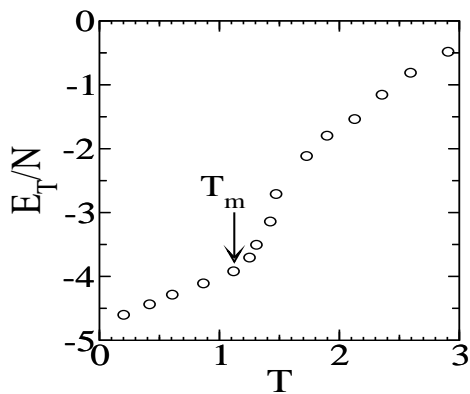}\\
  \caption{Energy as a function of temperature.
  The melting temperature is estimated as the inflexion
  point being around $T_m \approx 1.1 $}\label{EtxT}
\end{figure}

Our simulation is as follows. We consider the system as consisting
of an arrangement of particles of mass $m$, coupled by the
Lennard-Jones potential defined by the equation (\ref{equation1}).
The system is arranged in 4 layers with free boundary conditions in
all directions. The first layer is frozen in a regular arrangement
as in the $(001)$ surface of a Lennard-Jones bcc crystal in order to
maintain the whole structure as flat as possible. The tip is
simulated as a single particle of mass $m$, attached to three
springs of elastic constant $k_x=\epsilon_{ss}/\sigma_{ss}^2$
,$k_y=\epsilon_{ss}/\sigma_{ss}^2$ and
$k_z=\epsilon_{ss}/\sigma_{ss}^2$ as shown schematically in figure
\ref{FFMesqSim}. With the tip close to the surface we thermalize the
system at temperature $T$. After thermalization the tip is pushed in
a direction parallel to the surface at constant velocity $v_{0x}=v_L
\cos(\theta)$, $v_{0y}=v_L \sin(\theta)$, $v_{0z}=0$,
$v_L=v\sqrt{(\epsilon_{ss}/m)}$. Here $\theta$ is defined relatively
to the $x$ direction and $v$ is changed to obtain different
velocities. For each simulation the distance between the tip and the
surface is fixed at the beginning of the process, so that we can
control the perpendicular force on the tip. By measuring the size
variation of the springs we can calculate the laterals, $f_x$, $f_y$
forces and the perpendicular, $f_z$ force on the tip. This forces
are measured in units of, $\epsilon_{ss}/\sigma_{ss}$. The velocity,
position, energy and forces are stored at each time step for further
analysis. Before we start the simulation we have to estimate the
melting temperature $T_m$ of the system. Figure \ref{EtxT} shows the
total energy (per particle) as a function of temperature. The
melting temperature is estimated as the inflection point of the
curve. We find $T_m \approx 1.1$ in accordance with earlier
calculations \cite{pablo1,pablo2,flavio1}. Based in this result our
simulations will be performed with $T<T_m$ and the temperature will
be specified in each results.
%
%
\vspace{0.0cm}
%
We have simulated the system for several velocities, temperatures,
initial distance of the tip to the substrate or equivalently, normal
force in the tip and dragging angles relative to the surface. We are
mainly interested in studding the effects of the velocity, the
normal force and dragging angles relative to the surface in the
friction force and coefficient of friction.
\subsection{Dependence with velocity and normal force}
\begin{figure}[htbp]
\begin{center}
\includegraphics[width=8.50cm,height=9.5cm]{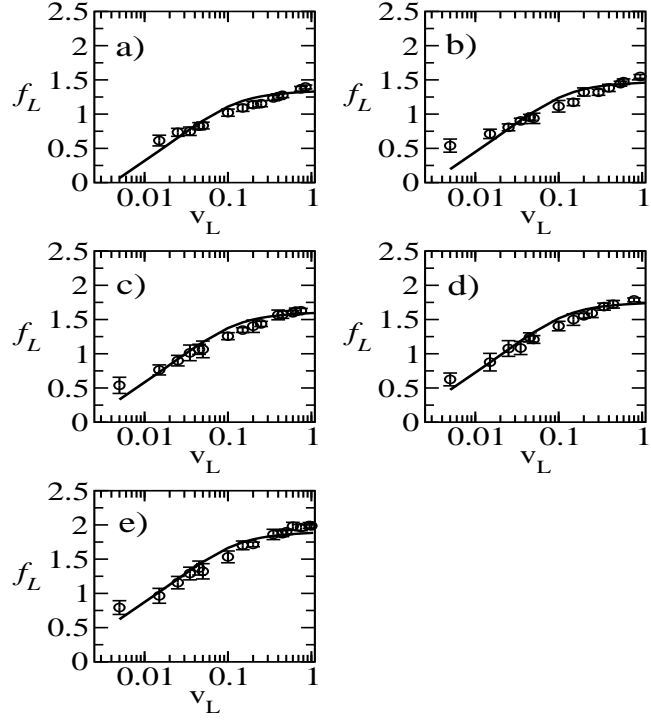}
\end{center}\vspace*{-0.50cm}
\caption{Friction force as a function of the velocity for some
selected values of the normal force. From a) to e) we have
respectively $<f_z>\sim 0.95$, $1.54$, $2.25$, $2.75$ and $2.50$.
The points are the results of the simulations and the lines are from
our analytical approach.}\label{FL1-vT05N0304050607Wplot}
\end{figure}

In this section we show the results of our simulations for several
values of the relative velocity  tip-surface and for five different
values of the normal force. In all the MD simulations of this
subsection we use $\theta=0$, $T=0.5$. In the figure
(\ref{FL1-vT05N0304050607Wplot}) we show the simulation (points) and
the theoretical (full lines) results for the friction force as a
function of the velocity for some normal forces. Note that our
theoretical result, equation (\ref{latforce}), is in very good
agreement with the results of the MD simulations, only adjusting
appropriate values for the critical position of jumps, $q_{zc}$ (See
section \ref{FirstPricMod}).
%
\vspace*{0.3cm}
%
\vspace* {0.0cm}
\begin{figure}[htbp]
\begin{center}
\includegraphics[width=5.0cm,height=5.0cm]{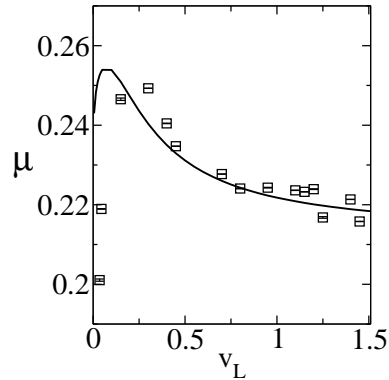}
\end{center}\vspace*{0.0cm}
\caption{Averaged friction coefficient as a function of the
velocity, calculated from the figure \ref{FL1-vT05N0304050607Wplot}.
The points are the results of the simulations and the line is the
analytical result.} \label{mi-vT05N0304050607Wplot}
\end{figure}
%
In the figure (\ref{mi-vT05N0304050607Wplot}) we show the results of
the simulation (points) and the theoretical results (full line) for
the averaged friction coefficient as function of the velocity. These
results are in good agreement with the simulation, indicating that
our model is consistent with the simulation.
%
%
\vspace* {0.0cm}
\begin{figure}[htbp]
\begin{center}
\includegraphics[width=8.0cm,height=8.0cm]{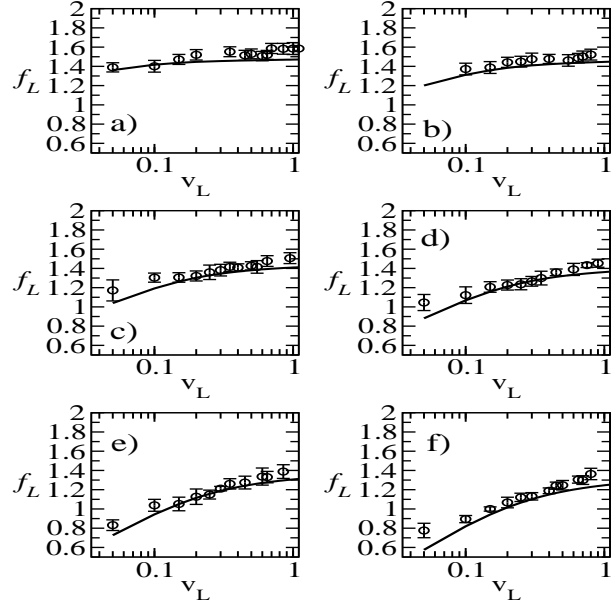}
\end{center}\vspace*{0.0cm}
\caption{Friction force as a function of the velocity for some
selected temperatures. The results of the simulations are shown as
circles the lines are from our analytical approach. The results from
a) to f) are respectively to $T=0.1$, $0.2$, $0.3$, $0.4$, $0.5$ and
$0.6$. }\label{Fl-vN3T010203040506}
\end{figure}
%
%
\vspace*{0.0cm}
\begin{figure}[htbp]
\begin{center}
\includegraphics[width=6.0cm,height=5.50cm]{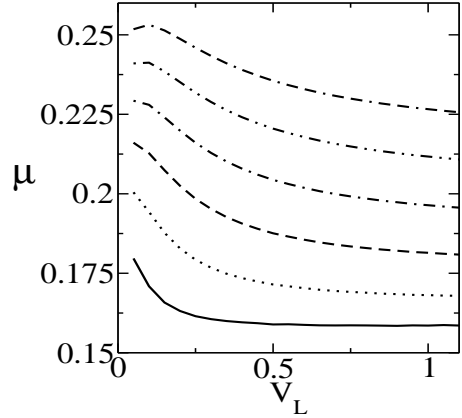}
\end{center}\vspace*{-0.50cm}
\caption{Averaged friction coefficient as a function of the velocity
for some selected temperatures. Here, the lines are for the
analytical results and from bottom to top corresponds respectively
to $T=0.1$, $T=0.2$, $T=0.3$, $T=0.4$, $T=0.5$ and
$T=0.6$.}\label{Coefricxv-TeoN3T0105}
\end{figure}
%
%
\subsection{Dependence with Temperature and Velocity.}
In this section we present some simulation for the friction force
for several velocities, and different values of temperature,
($T=0.1$, $0.2$, $0.3$, $0.4$, $0.5$, $0.6$). In all the simulations
in this subsection we use $\theta=0$ and $<f_z> \sim 0.95$. In the
figure (\ref{Fl-vN3T010203040506}) are the results of the MD
simulations (points) and the theoretical result (full lines) for the
friction force. Note that our theoretical results are in excellent
agreement with the results of the MD simulation. In the figure
(\ref{Coefricxv-TeoN3T0105}) we show the analytical results for the
averaged friction coefficient as a function of the velocity. In this
figure, the lines from bottom to top corresponds respectively to
$T=0.1$, $T=0.2$, $T=0.3$, $T=0.4$,$T=0.5$, $T=0.6$.

In figures (\ref{mi-vT05N0304050607Wplot}) and
(\ref{Coefricxv-TeoN3T0105}), it is shown the behavior of the
friction coefficient as a function of the relative velocity. We note
that initially the friction coefficient increases with the velocity
until it reaches a maximum and then starts to diminish. The initial
increasing and the maximum value for the coefficient of friction can
be related to the fact that when the velocity increases the tip
executes jumps more and more energetic, and a high friction mode
associated with single slips appears as indicated by Nakamura et.
al. \cite{nakamura:235415}. As the velocity pass though a certain
limit a low friction mode with double slips appears leading to the
observed reduction of the friction coefficient.
%
%
\vspace* {0.0cm}
\begin{figure}[htbp]
\begin{center}
\includegraphics[width=8.50cm,height=9.50cm]{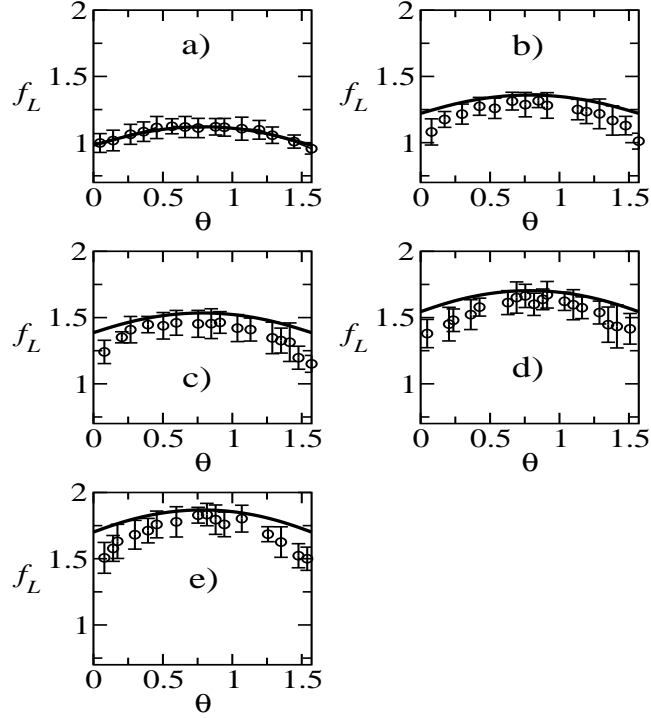}
\end{center}\vspace*{0.0cm}
\caption{Friction force as function of the angle, $\theta$, for some
selected normal forces. From a) to e) we have respectively
$<f_z>\sim 0.95$, $1.54$, $2.25$, $2.75$, $2.50$. The points are the
results of the simulations and the lines are our theoretical
results. }\label{FlxthetaT05N0304050607}
\end{figure}
%
%
\subsection{Dependence with Normal Force and Force Pushing Angle.}
Below we present the simulations for five different values of the
normal force and several pulling angles, $0\leq\theta\leq\pi/2$,
between the direction $x$ and the force pulling the cantilever. In
all the MD simulations in this section we use $v_l=0.05$, $T=0.5$.
%
%
\vspace*{0.5cm}
\begin{figure}[htbp]
\begin{center}
\includegraphics[width=6.0cm,height=5.0cm]{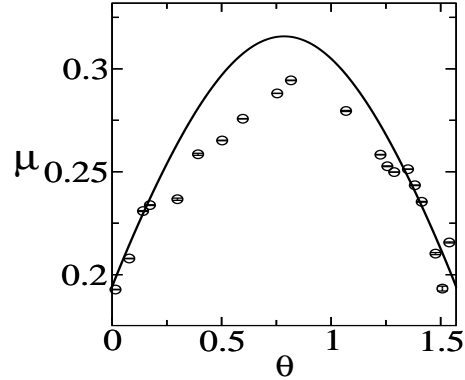}
\end{center}\vspace*{-0.50cm}
\caption{Averaged friction coefficient as a function of the angle,
$\theta$, calculated from the figure \ref{FlxthetaT05N0304050607}.
The points are the results of the simulations and the line is the
analytical result.} \label{mixtheta-TeoMD}
\end{figure}
In the figure (\ref{FlxthetaT05N0304050607}) and
(\ref{mixtheta-TeoMD}) we show our results of the simulation
(points) and the analytical calculations (full lines) for the
friction force and the averaged friction coefficient as a function
of the angle, $\theta$. Note that our theoretical result, equation
(\ref{latforce}), is in very good agreement with the results of the
MD simulations.
%
%
%
\section{Conclusion and Comments\label{concluComment}}
In this paper we have proposed a tri-dimensional model to describe
friction force. Our approach generalize the one-dimensional friction
force model proposed in earlier works \cite{Riedo,Sang}. In our
model we include the influence of the normal force which allow us to
calculate the friction coefficient for first principles. We apply
our model to calculate the friction force and friction coefficient
for a tip-surface interaction. The interaction between the tip and
the surface is represented by a Lennard-Jones potential and the
surface represented by a potential that represent a (001) surface
symmetry and the friction force obtained are in good agreement with
the experimental results presentes by Riedo\cite{Riedo}. With the
intend to test our model we performed classical molecular-dynamics
simulation for a particle attached in a spring and pushed over a
BCC(001) surface crystal. For this purpose we performed the
simulation for several velocities, temperatures, normal forces and
dragging angle. We have found that the friction forces and the
friction coefficient are in good agreement with our molecular
dynamic results, proving the effectiveness of our model.

\subsection{Acknowledgments}
This work was supported by CNPq. We are grateful to Ms. B. A. Soares
for suggestions and comments.

%

\bibliographystyle{unsrt} 
\bibliography{referMDVe}

\begin{thebibliography}{10}

\bibitem{Nanoscience}
E.~Meyer, R.M. Overney, K.~Dransfeld, and T.~Gyalog.
\newblock {\em Nanoscience - Friction and Rheology on the Nanometer Scale}.
\newblock World Scientific, 1998.

\bibitem{HBMicroNanoTribo}
Bharat Bhushan.
\newblock {\em Handbook of Micro/Nano Tribology}.
\newblock CRC Press., second edition, 1999.

\bibitem{Gnecco}
E.~Gnecco, R.~Bennewitz, T.~Gyalog, Ch. Loppacher, M.~Bammerlin, E.~Meyer, and
  H.-J. G\"untherodt.
\newblock Velocity dependence of atomic friction.
\newblock {\em Phys. Rev. Lett.}, 84:1172, 2000.

\bibitem{Riedo}
E.~Riedo, E.~Gnecco, R.~Bennewitz, E.~Meyer, and H.~Brune.
\newblock Interaction potential and hopping dynamics governing sliding
  friction.
\newblock {\em Phys. Rev. Lett.}, 91:084502, 2003.

\bibitem{EGneccoWear254}
E.~Gnecco, R.~Bennewitz, A.~Socoliuc, and E.~Meyer.
\newblock Friction and wear on the atomic scale.
\newblock {\em Wear}, 254:859--862, 2003.

\bibitem{chen:236102}
Jinyu Chen, Imma Ratera, Jeong~Young Park, and Miquel Salmeron.
\newblock Velocity dependence of friction and hydrogen bonding effects.
\newblock {\em Physical Review Letters}, 96(23):236102, 2006.

\bibitem{EuiSungYoon}
Eui-Sung Yoon, R.~Arvind Singh, Hyun-Jin Oh, and Hosung Kong.
\newblock The effect of contact area on nano/micro-scale friction.
\newblock {\em Wear}, 259:1424--1431, 2005.

\bibitem{Sang}
Y.~Sang, M.~Dub\'e, and M.~Grant.
\newblock Thermal effects on atomic friction.
\newblock {\em Phys. Rev. Lett.}, 87:174301, 2001.

\bibitem{PerssonWear}
B.~N.~J. Persson, O.~Albohr, F.~Mancosu, V.~Peveri, V.~N. Samoilov, and I.~M.
  Sivebaek.
\newblock On the nature of the static friction, kinetic friction and creep.
\newblock {\em Wear}, 254(9):835--851, 2003.

\bibitem{Mate}
C.~Mathew Mate, Gary~M. McClelland, Ragnar Erlandsson, and Shirley Chiang.
\newblock Atomic-scale friction of a tungsten tip on a graphite surface.
\newblock {\em Phys. Rev. Lett.}, 59(17):1942--1945, Oct 1987.

\bibitem{ASocoliuc}
A.~Socoliuc, R.~Bennewitz, E.~Gnecco, and E.~Meyer.
\newblock Transition from stick-slip to continuous sliding in atomic friction:
  Entering a new regime of ultralow friction.
\newblock {\em Phys. Rev. Lett.}, 92:134301, 2004.

\bibitem{PRB62}
H.~H\"olscher, W.~Allers, U.~D. Schwarz, A.~Schwarz, and R.~Wiesendanger.
\newblock Interpretation of \char16{}true atomic resolution\char17{} images of
  graphite (0001) in noncontact atomic force microscopy.
\newblock {\em Phys. Rev. B}, 62(11):6967--6970, Sep 2000.

\bibitem{Steele}
W.~Steele.
\newblock The physical interaction of gases with crystalline solids.
\newblock {\em Surf. Sci.}, 36:317, 1973.

\bibitem{RadiasBJP}
R.A. Dias, M.~Rapini, P.Z. Coura, and B.V. Costa.
\newblock Temperature dependent molecular dynamic simulation of friction.
\newblock {\em Brazilian Journal of Physics}, 36(3A):741--745, September 2006.

\bibitem{lantz:970}
M.~A. Lantz, S.~J. O'Shea, A.~C.~F. Hoole, and M.~E. Welland.
\newblock Lateral stiffness of the tip and tip-sample contact in frictional
  force microscopy.
\newblock {\em Applied Physics Letters}, 70(8):970--972, 1997.

\bibitem{carpick:1548}
R.~W. Carpick, D.~F. Ogletree, and M.~Salmeron.
\newblock Lateral stiffness: A new nanomechanical measurement for the
  determination of shear strengths with friction force microscopy.
\newblock {\em Applied Physics Letters}, 70(12):1548--1550, 1997.

\bibitem{PhysRevB.52.14976}
U.~D. Schwarz, W.~Allers, G.~Gensterblum, and R.~Wiesendanger.
\newblock Low-load friction behavior of epitaxial $c_{60}$ monolayers under
  hertzian contact.
\newblock {\em Phys. Rev. B}, 52(20):14976--14984, Nov 1995.

\bibitem{FogdenWhite}
A.~Fogden and Lee~R. White.
\newblock Contact elasticity in the presence of capillary condensation : I. the
  nonadhesive hertz problem.
\newblock {\em Journal of Colloid and Interface Science}, 138:414--430, 1990.

\bibitem{SangminJeon}
S.~Jeon, T.~Thundat, and Yehuda Braiman.
\newblock Effect of normal vibration on friction in the atomic force microscopy
  experiment.
\newblock {\em Appl. Phys. Lett.}, 88:214102, 2006.

\bibitem{PRL93}
Makoto Ashino, Alexander Schwarz, Timo Behnke, and Roland Wiesendanger.
\newblock Atomic-resolution dynamic force microscopy and spectroscopy of a
  single-walled carbon nanotube: Characterization of interatomic van der waals
  forces.
\newblock {\em Physical Review Letters}, 93(13):136101, 2004.

\bibitem{PerssonNitzan}
B.N.J. Person and A.~Nitzan.
\newblock Linear sliding friction: on the origin of the microscopic friction
  for xe on silver.
\newblock {\em Surface Science}, 367:261--275, 1996.

\bibitem{Liebsch}
A.~Liebsch, S.~Gonçalves, and M.~Kiwi.
\newblock Electronic versus phononic friction of xenon on silver.
\newblock {\em Phys. Rev. B}, 60:5034, 1999.

\bibitem{Tomassone}
M.S. Tomassone, J.B. Sokoloff, A.~Widom, and J.~Krim.
\newblock Dominance of phonon friction for a xenon film on a silver (111)
  surface.
\newblock {\em Phys. Rev. Lett.}, 79:4798, 1997.

\bibitem{Allen}
M.P. Allen and D.J. Tildesley.
\newblock {\em Computer Simulation of Liquids}.
\newblock Oxford Scince Publications, 1992.

\bibitem{Beeman}
D.~Bemman.
\newblock Some multistep method for use in melecular dynamic calculations.
\newblock {\em J. Comput. Phys.}, 20:130--139, 1976.

\bibitem{Berendsen}
H.~J.~C. Berendsen and W.~F. Gunsteren.
\newblock {\em Pratical Algorithms for Dynamic Simulations}, pages 43--65.

\bibitem{pablo1}
P.Z. Coura, O.N. Mesquita, and B.V. Costa.
\newblock Molecular-dynamics simulation of directional growth of binary
  mixtures.
\newblock {\em Phys. Rev. B}, 59:3408, 1999.

\bibitem{pablo2}
P.Z. Coura, O.N. Mesquita, and B.V. Costa.
\newblock Molecular dynamics simulation of zone melting.
\newblock {\em Int. J. Mod. Phys. C}, 9(6):857--860, 1998.

\bibitem{flavio1}
F.J. Resende and B.V. Costa.
\newblock Molecular-dynamics study of the diffusion coefficient on a crystal
  surface.
\newblock {\em Phys. Rev. B}, 61:12697, 2000.

\bibitem{flavio2}
F.J. Resende and B.V. Costa.
\newblock Molecular dynamics study of copper cluster deposition on a (010)
  surface.
\newblock {\em Surface Science}, 481:54, 2001.

\bibitem{Rapaport}
D.~C. Rapaport.
\newblock {\em The Art of Molecular Dynamics Simulation}.
\newblock Cambridge University Press, 2000.

\bibitem{nakamura:235415}
Jun Nakamura, Shinya Wakunami, and Akiko Natori.
\newblock Double-slip mechanism in atomic-scale friction: Tomlinson model at
  finite temperatures.
\newblock {\em Physical Review B (Condensed Matter and Materials Physics)},
  72(23):235415, 2005.

\end{thebibliography}
\end{document}